\RequirePackage{arydshln}
\documentclass[floatfix,aps,twocolumn,nofootinbib,superscriptaddress,preprintnumbers,pra,10pt]{revtex4-1}

\usepackage[utf8]{inputenc}
\usepackage{float}
\usepackage{colortbl}
\usepackage{amsmath,amssymb,amsfonts,bm,bbm,slashed}
\usepackage{dsfont} 
\usepackage{hyperref}
\usepackage{graphicx}
\usepackage{enumitem}
\usepackage{arydshln}
\usepackage{mathtools}
\usepackage{bbold}
\usepackage{braket}
\usepackage{multirow}
\usepackage{soul}
\usepackage{xcolor}
\usepackage{soul}
\usepackage{lipsum}
\usepackage{subfig}

\makeatletter
\g@addto@macro\bfseries{\boldmath}
\makeatother

\newcommand{\be}{\begin{align}}
\newcommand{\ee}{\end{align}}
\newcommand{\bea}{\begin{eqnarray}}
\newcommand{\eea}{\end{eqnarray}}
\newcommand{\ba}{\begin{align}}
\newcommand{\ea} {\end{align}}

\newcommand{\GeV}{\text{GeV}}

\newcommand{\GF}{G_F}
\newcommand{\Rpi}{R_\pi}
\newcommand{\cB}{\mathcal{B}}
\renewcommand{\O}{\mathcal{O}}

\renewcommand{\Im}{\text{Im}}
\newcommand{\labdt}{\lambda^{(d)}_t}
\newcommand{\labdu}{\lambda^{(d)}_u}
\newcommand{\labdc}{\lambda^{(d)}_c}
\newcommand{\labst}{\lambda^{(s)}_t}

\newcommand{\labsc}{\lambda^{(s)}_c}
\newcommand{\gsim}{\lower.7ex\hbox{$\;\stackrel{\textstyle>}{\sim}\;$}}
\newcommand{\lsim}{\lower.7ex\hbox{$\;\stackrel{\textstyle<}{\sim}\;$}}

\def\eqs#1#2{{Eqs.~(\ref{#1})--(\ref{#2})}}
\def\fig#1{{Fig.~\ref{#1}}}

\def\Table#1{{Table~\ref{#1}}}

\def\sec#1{{Sect.~\ref{#1}}}

\def\app#1{{Appendix~\ref{#1}}}

\newcommand{\abs}[1]{\left| #1 \right|}

\allowdisplaybreaks

\begin{document}
%%%%%%%%%%%%%%%%%%%%%%%%%%%%%%%%%%%%%%%%%%%%%%%%%%%%%%%%%%%%

\preprint{SI-HEP-2021-03, P3H-21-005, TUM-HEP-1315/21, ZH-TH 2/21}

\title{The LFU Ratio \texorpdfstring{$\Rpi$}{R(pi)} in the Standard Model and Beyond}
 
 \author{Marzia Bordone}
 \email{marzia.bordone@to.infn.it}
\affiliation{Theoretische Physik 1, Naturwissenschaftlich-Technische Fakult\"at, Universit\"at Siegen, Walter-Flex-Stra{\ss}e 3, D-57068 Siegen, Germany \&
Dipartimento di Fisica, Universit\`a di Torino \& INFN, Sezione di Torino, I-10125 Torino, Italy}
\author{Claudia Cornella}
\email{claudia.cornella@physik.uzh.ch}
\affiliation{Physik-Institut, Universit\"at Z\"urich, CH-8057 Z\"urich, Switzerland}
\author{Gino Isidori}
\email{isidori@physik.uzh.ch}
\affiliation{Physik-Institut, Universit\"at Z\"urich, CH-8057 Z\"urich, Switzerland}
\author{Matthias K\"{o}nig}
\email{matthias.koenig@tum.de}
\affiliation{Physik Department T31, Technische Universit\"at M\"unchen, James-Franck-Stra\ss{}e 1, D-85748 Garching, Germany}

\begin{abstract}
\vspace{5mm}
We discuss the possibility of performing precise tests of $\mu/e$ universality in $B \to\pi \ell^+\ell^-$ decays. We show that in wide regions of the dilepton invariant mass spectrum the ratio between muonic and electronic decay widths can be predicted with high accuracy, both within and beyond the Standard Model. We present numerical expressions which can be used to extract precise information on short-distance dynamics if a deviation from universality is observed in the data.
\vspace{3mm}
\end{abstract}

\maketitle

%%%%%%%%%%%%%%%%%%%%%%%%%%%%%%%%%%%%%%%%%%%%%%%%%%
\section{Introduction}\label{sec:intro}
%%%%%%%%%%%%%%%%%%%%%%%%%%%%%%%%%%%%%%%%%%%%%%%%%%

The experimental measurements of the $\mu/e$ universality ratios $R_K$ and $R_{K^*}$ in $B \to K^{(*)} \ell^+\ell^-$ decays~\cite{Aaij:2014ora,Aaij:2017vbb,Aaij:2019wad,Abdesselam:2019wac}  
indicate a violation of Lepton Flavor Universality (LFU) of about $20\%$ in the decay rates, well above the Standard Model (SM) expectation~\cite{Hiller:2003js,Bordone:2016gaq}. The statistical significance of each measurement does not exceed the $3\sigma$ level. However, as pointed out first in \cite{Hiller:2014yaa}, these results are consistent with the tension between data and SM predictions in the $B \to K^{*} \ell^+\ell^-$ differential distribution~\cite{Aaij:2020nrf,Aaij:2015oid}, as well as with the suppression of $\cB(\bar B_s\to \mu^+\mu^-)$~\cite{LHCb:2020zud} compared to the SM expectation~\cite{Beneke:2017vpq,Beneke:2019slt}. When combined, the $b\to s\ell \ell$ data points towards a non-standard phenomenon of short-distance origin, with a statistical significance exceeding $4\sigma$ (see Refs.~\cite{Alguero:2019ptt,Aebischer:2019mlg,Ciuchini:2020gvn} for recent combined analyses).

Interestingly enough, an independent indication of  LFU violation 
occurs in $\bar B \to D^{(*)} \ell \nu$ decays, when comparing $\tau$ and light-lepton modes~\cite{Aaij:2015yra,Lees:2013uzd,Hirose:2016wfn,Aaij:2017deq,Abdesselam:2019dgh}.
These two hints of LFU violation, generically referred to as $B$-physics anomalies, can be addressed by a combined description within an effective theory approach to physics beyond the SM based on two main hypotheses: \textit{i})~the new dynamics affect predominantly semi-leptonic operators, and \textit{ii})~they couple in a non-universal way to the different fermion generations~\cite{Bhattacharya:2014wla, Alonso:2015sja, Greljo:2015mma, Calibbi:2015kma, Barbieri:2015yvd,Buttazzo:2017ixm}. In particular, the new dynamics should have dominant couplings to third-generation fermions and smaller, but non-negligible, couplings to  second-generation fermions. This non-trivial flavor structure resembles the hierarchies observed in the SM Yukawa couplings, opening the possibility of a common origin of $B$-physics anomalies and flavor hierarchies, as hypothesized in Refs.~\cite{Bordone:2017bld,Fuentes-Martin:2020pww}.

In order to shed light on this phenomenon, it would be very important to establish evidence of the same underlying non-standard dynamics in different channels. In particular, most frameworks addressing $B$-physics anomalies predict a sizable violation of $\mu/e$ universality also in processes based on the $b\to d \ell \ell$ transitions.  The connection between $b\to s$ and $b\to d$ flavor-changing neutral-current (FCNC) amplitudes is a firm prediction of all models based on a minimally broken $U(2)^5$ flavor symmetry~\cite{Barbieri:2011ci,Blankenburg:2012nx, Barbieri:2012uh}, such as the ones proposed in Refs.~\cite{Bordone:2017bld,Fuentes-Martin:2020pww}, and the wider class models discussed in Refs.~\cite{Greljo:2018tuh,DiLuzio:2018zxy,Cornella:2019hct,Fuentes-Martin:2019ign}~(the so-called non-universal 4321 models). This connection has been discussed in general terms in Ref.~\cite{Fuentes-Martin:2019mun}, where it has been shown that the $U(2)^5$  symmetry implies an identical relative breaking of LFU in $b\to s \ell \ell$ and $b\to d \ell \ell$ amplitudes at the short-distance level~\cite{Fuentes-Martin:2019mun}.

On general grounds, testing the SM precisely in $b\to d \ell \ell$-mediated processes is challenging for two reasons. 
First, the rates are both loop- and CKM-suppressed, making the decay extremely rare. Second, hadronic intermediate states introduce sizable long-distance contributions, which are difficult to predict. A promising channel in this regard is $B \to \pi \ell^+\ell^-$: about $20$ events of $B \to \pi \mu^+\mu^-$ have already been observed by LHCb in the LHC run-I~\cite{Aaij:2015nea}, and a significantly larger sample can be anticipated from run-II data. Recent theoretical studies of $B \to \pi \ell^+\ell^-$ decays, analyzing the relative impact of short- and long-distance contributions within the SM, have been presented in Refs.~\cite{Ali:2013zfa, Hou:2014dza, Hambrock:2015wka, Khodjamirian:2017fxg}. A key point to notice is that long-distance contributions cannot induce violations of LFU. Hence, as in  $B \to K \ell^+\ell^-$ decays, ratios of the type~\cite{Hiller:2003js}
\begin{align}
\Rpi[q^2_\text{min}, q^2_\text{max}] =\frac{ \int_{q^2_\text{min}}^{q^2_\text{max}} dq^2\frac{d\mathcal{B}}{dq^2}(B^+\to\pi^+\mu^+\mu^-)}{ \int_{q^2_\text{min}}^{q^2_\text{max}}dq^2\frac{d\mathcal{B}}{dq^2}(B^+\to\pi^+e^+e^-)}\,,
\label{eq:Rpidef}
\end{align}
with $q^2=m^2_{\ell \ell}$ being the dilepton invariant mass, are expected to provide powerful tests of LFU violation of short-distance origin.
The purpose of this paper is to precisely estimate the sensitivity of these ratios to short-distance dynamics.

While long-distance contributions cannot induce a violation of LFU, they can {\em dilute} a possible LFU-violating contribution of short-distance origin in kinematical regions where they are dominant. Our main goal is therefore to identify the regions of the dilepton invariant mass spectrum where the decay rate is dominated by short-distance dynamics and to estimate the sensitivity to LFU-violating amplitudes in those regions. We will do so using a general parameterization of the problematic long-distance contributions induced by light-quark and charmonium vector resonances decaying into $\ell^+\ell^-$ pairs. We employ a data-driven approach, following the strategy originally proposed in Ref.~\cite{Kruger:1996cv}, and further developed in Refs.~\cite{Khodjamirian:2012rm,Lyon:2014hpa,Blake:2017fyh,Bobeth:2017vxj,Cornella:2020aoq} in the context of $b\to s \ell \ell$ transitions. As we shall show, in a wide range of $q^2$ values, $\Rpi$ can allow us to extract precise information on LFU-violating dynamics.

%---------------------------------------------------------------------------------------------------------------------------------------------------------------------------------------------------------------------
\section{Theoretical description of \texorpdfstring{$B\to \pi \ell^+\ell^-$}{B->Pi l l} decays}
\label{sec:2}
%---------------------------------------------------------------------------------------------------------------------------------------------------------------------------------------------------------------------

The starting point to describe $B\to \pi \ell^+\ell^-$ decays 
is the  $b\to d \ell^+\ell^-$ effective Hamiltonian.
Since we are interested in the comparison between $B\to \pi$  and $B\to K$ 
modes, we keep $d$ as a generic label for down-type quarks whenever possible, and 
we generically denote the light final-state meson by $P$.

The effective Hamiltonian is
\begin{align}
\begin{aligned}
&& \mathcal{H}(b\to d\ell^+\ell^-)= \frac{4\GF}{\sqrt{2}} \bigg\{\labdc (\mathcal{C}_1\mathcal{O}_1^c+C_2\mathcal{O}_2^c) \\
&& \qquad +\labdu (\mathcal{C}_1\mathcal{O}_1^u+\mathcal{C}_2\mathcal{O}_2^u)-\labdt\sum_{i=3}^{10}C_i\mathcal{O}_i\bigg\}\,,
\end{aligned}
\end{align}
where $\lambda^{(d)}_i=V^*_{id}V_{ib}$ and, due to CKM unitarity, $\labdt+\labdu+\labdc=0$. 
The leading FCNC operators are defined as 
\begin{align}
\begin{aligned}
\O_{7}=& \ \frac{e}{(4\pi)^2}(\bar{d} \sigma^{\mu\nu}  (m_b P_{R} + m_{d} P_{L})  b)F_{\mu\nu} \,, \\
\O_{9}=& \ \frac{e^2}{(4\pi)^2}(\bar{d}_L\gamma_\mu b_L)(\bar\ell \gamma^\mu\ell)  \,,  \\
\O_{10}=& \ \frac{e^2}{(4\pi)^2}(\bar{d}_L\gamma_\mu b_L)(\bar\ell\gamma^\mu\gamma_5\ell) \, ,
\end{aligned}
\label{eq:operators_1}
\end{align}
while the leading four-quark operators read ($q=u,c$):
\begin{align}
\begin{aligned}
\O_1^q=& \ (\bar{d}_L^{\alpha} \gamma_\mu  q_L^\beta)(\bar{q}_L^\beta \gamma^\mu b_L^\alpha) \,, \\
\O_2^q=& \ (\bar{d}_L \gamma_\mu q_L)(\bar{q}_L\gamma^\mu b_L) \, .
\end{aligned}
 \label{eq:operators_2}
 \end{align}
 The NNLO expressions for the  Wilson coefficients of the operators in \eqs{eq:operators_1}{eq:operators_2} can be found in Ref.~\cite{Gorbahn:2004my}.
For reference, with this normalization we use $\mathcal{C}_{9}^{\rm SM} \approx 4.1$, $\mathcal{C}_{10}^{\rm SM} \approx -4.3$ and $\mathcal{C}_7^\text{SM}\approx -0.29$.

The hadronic matrix elements of quark bilinears in $\bar{B}\to P$ decays, where $P$ is a pseudoscalar meson containing a $d$ quark, can be parametrized as
\begin{align}
\begin{aligned}
\langle P(k)|\bar{d} \gamma^\mu b |\bar{B}(p)\rangle =& \left[(p+k)^\mu-\frac{m_B^2-m_P^2}{q^2}q^\mu\right]f_+(q^2)  \\
&+\frac{m_B^2-m_P^2}{q^2}q^\mu\,f_0(q^2)\,, \\
\! \langle P(k)|\bar{d} \sigma^{\mu\nu}q_\nu b |\bar{B}(p)\rangle =& \frac{i}{m_{B}+m_P}\Big[2 q^2 p^\mu \\
& -(m_{B}^2-m_P^2+q^2)q^\mu \Big]f_{T}(q^2)\,,
\end{aligned}
 \end{align}
where $q^\mu = p^\mu-k^\mu$ is the momentum transfer. The form factors $f_+(q^2)$, $f_0(q^2)$ and $f_T(q^2)$ can be extracted from Lattice QCD and Light-Cone Sum Rules (LCSR) techniques. Concerning $B\to \pi $ decays, we use the results of Ref.~\cite{Gubernari:2018wyi}, which combines 
LCSR estimates with Lattice QCD calculations from Ref.~\cite{Lattice:2015tia}. 

Taking into account only the contribution of operators with non-vanishing tree-level matrix elements, the differential decay width for the semi-leptonic $\bar{B}\to P\ell^+\ell^-$ decay reads:
\begin{align}
\begin{aligned}
 \frac{d\Gamma}{dq^2}=& \frac{\alpha_\text{em}^2 G_F^2 \left|\lambda_{t}^{(d)}\right|^2}{1024 \pi^5 m_B^3} \beta(q^2 )\lambda^{1/2}(q^2)\, \times \\
& \left\{\lambda(q^2)\left[1-\frac{\beta^2(q^2)}{3}\right]\times \right. \\
&\qquad \left| \mathcal{C}_9^\ell f_+(q^2) +2 \frac{m_b+m_d}{m_B+m_P}\mathcal{C}^\ell_7f_T(q^2)\right|^2  \\
&+\frac{2}{3}\lambda(q^2) \beta^2(q^2) \, |\mathcal{C}^\ell_{10} f_+(q^2)|^2 \\
& \left. +4 m_\ell^2 \frac{(m_B^2-m_P^2)^2}{q^2 }|\mathcal{C}_{10}^\ell f_0(q^2)|^2\right\}\,,
\end{aligned}
\end{align}
where
\begin{align}
\begin{aligned}
\lambda (q^2) &= (m_{B}^2-m_P^2-q^2)^2- 4 m_P^2 q^2\,, \\
\beta(q^2)&=\sqrt{1-4m_\ell^2/q^2}\,.
\end{aligned}
\end{align}
The superscript $\ell$ in the Wilson coefficients denotes the lepton flavor in the final state. We recall that, in the SM, the Wilson coefficients are universal for the three lepton generations, i.e. $\mathcal{C}_i^{\tau}=\mathcal{C}_i^{\mu}=\mathcal{C}_i^{e} $, where the index $i$ runs over all the possible Wilson coefficients.

\subsection{Non-local contributions}
In order to account for the contribution of the four-quark operators and describe the 
$B\to P\ell^+\ell^-$ spectrum also in the resonance region, we modify the Wilson coefficient $\mathcal{C}_9^\ell$ as follows~\cite{Khodjamirian:2012rm,Cornella:2020aoq}:
\begin{align}
\begin{aligned}
\mathcal{C}_9^\ell \to \mathcal{C}_9^{\ell,\text{eff}}(q^2) =\,& \mathcal{C}_9^\ell -  \frac{\labdc}{\labdt} 
Y_{c\bar c}(q^2) -  \frac{\labdu}{\labdt} Y_{u\bar u}(q^2)  \\
& +Y_{d\bar d}(q^2)+Y_{s\bar s}(q^2)~, \\
 \equiv\,& \mathcal{C}_9^\ell -  \frac{\labdc}{\labdt} 
Y_{c\bar c}(q^2) + Y_{\rm light}(q^2)~.
\end{aligned}
\end{align}
Here $Y_{q\bar q}(q^2)$ denotes the non-local contribution due to intermediate hadronic states with $q\bar q$ valence quarks. For later convenience, we have grouped the contribution induced by light quarks into a single function $Y_{\rm light}(q^2)$. We express the $Y$ functions as
\begin{align}
\begin{aligned}
Y_{q\bar q}(q^2) &=  \frac{16\pi^2}{f_+(q^2)}\mathcal{H}_{q\bar q}(q^{2})\,, \\
Y_{\rm light}(q^2) &=  \frac{16\pi^2}{f_+(q^2)}\mathcal{H}_{\rm light}(q^{2})~.
\label{eq:Ydef}
\end{aligned}
\end{align}
In principle, $\mathcal{H}_{c\bar c}(q^{2})$ could be evaluated through the correlation function~\cite{Khodjamirian:2012rm}
\begin{align}
\begin{aligned}
 &i\int d^4x e^{iqx}\langle P(k)|T\left\{j_\mu^\text{em}(x), \sum_{i=1,2} \mathcal{C}_i^c\mathcal{O}_i^q(0) \right\}|\bar{B}(p)\rangle  \\
 & =\,[(p\cdot q)q_\mu-q^2 p_\mu]\mathcal{H}_{c\bar c}(q^2)\,,
\end{aligned}
\end{align}
with $j_\mu^\text{em}= \sum_{q=u,d,s,c,b}Q_q \bar{q}\gamma_\mu q$. 
A similar expression can be derived for $\mathcal{H}_{\rm light}(q^{2})$ in terms of the charmless operators.
In practice, we are unable to evaluate these expressions from first principles and we estimate them from data 
using dispersion relations~\cite{Khodjamirian:2012rm,Lyon:2014hpa,Blake:2017fyh,Bobeth:2017vxj,Cornella:2020aoq}.

\subsubsection{Estimate via dispersion relations}

In full generality, we can write a subtracted dispersion relation for $\mathcal{H}_{q\bar q}(q^{2})$
\begin{align}
\Delta Y_{q\bar q}(q^{2}) = \frac{16 \pi^{2}}{f_+(q^2)} \Delta \mathcal{H}_{q \bar q}(q^{2})\, ,
\end{align}
with
\begin{align}
\begin{aligned}
\Delta \mathcal{H}_{q \bar q}(q^{2})  &= \frac{q^2- q_{0}^2}{\pi} \int_{s_0}^\infty ds \frac{ \Im[\mathcal{H}_{q\bar q}(s)]}{(s-q_{0}^{2})(s-q^2)} \\
&\equiv \frac{q^2-q_{0}^2}{\pi}\int_{s_0}^\infty ds \frac{\rho_{q\bar q}(s)}{(s-q_{0}^{2})(s-q^2)}\,,
\end{aligned}
\end{align}
The function $\rho_{q\bar q}(s)$ is the spectral density for an intermediate hadronic state with valence quarks $q\bar q$ and invariant mass $s$, and the parameter $s_0$ is the energy threshold where the state can be created on-shell. The parameter $q_0^2$ is the subtraction point that we choose for the different $q\bar q$ states ($q_{0}^{2}<s_{0}$). 
%The spectral density is an infinite sum over all the possible hadronic states with the characteristics just listed. 

The leading contribution to $\rho_{q\bar q}(s)$ is provided by single-particle intermediate states. We can describe them as a sum of  Breit-Wigner distibutions:
\begin{align}
\begin{aligned}
\Delta \mathcal{H}_{q\bar q}^{\text{1P} } &= \sum_{V_j}\eta_j e^{i\delta_j}\frac{(q^2-q_0^2)}{(m_j^2-q_0^2)} A_j^\text{res}(q^2)\,, \\
 A_j^\text{res}(q^2) &= \frac{m_j \Gamma_j}{m_j^2-q^2-i m_j \Gamma_j} \,,
\end{aligned}
\end{align}
where the sum runs over all the possible vector states associated with the the $q\bar{q}$ valence quarks. The parameters $\eta_j$ and $\delta_j$ have to be determined from data. 
For the charmonium resonances, which have %a clear valence structure ($c\bar c$  only) 
a high invariant mass, we use dispersion relations subtracted at $q^2=0$, yielding
\begin{align}
\Delta \mathcal{H}^\mathrm{1P}_{c\bar c} = \sum_{V_j =J/\psi, \psi(2S),\ldots}\eta_j e^{i\delta_j} \frac{q^2}{m^2_{\psi_j}}  A_j^\mathrm{res}(q^2)\,.
\end{align}
For the light resonances we use unsubtracted dispersion relations, which is equivalent to assuming a vanishing long-distance contribution from light quarks in the large-$q^2$ limit. In this case we do not separate the various flavors explicitly, obtaining
\begin{align}
\mathcal{H}^\mathrm{1P}_{\rm light} = \sum_{V_j= \rho, \omega, \phi }\eta_j e^{i\delta_j} A_j^\mathrm{res}(q^2)\,.
\end{align}

\begin{table*}[t]
\begin{center}
\renewcommand{\arraystretch}{1.3} 
\begin{tabular}{|c|| c| c|c| c|}
\hline
$V$ & $m_{V} \,[ \text{MeV}]$ & $ \Gamma_{V}  \, [\text{MeV}] $ & $\mathcal{B}(B\to \pi V)$ &  $\mathcal{B}(V \to e^{+} e^{-})$\\
\hline \hline
$\rho$ &   $ 775.25\pm 0.26 \, $ &  $ 147.8\pm 0.9 $ &  $(8.3\pm 1.2)\times 10^{-6}$ & $(4.72\pm 0.05)\times 10^{-5}$ \\
\hline
$\omega$ &   $782.65\pm 0.12 $ & $8.49\pm 0.08 $ & $(6.9\pm 0.5)\times 10^{-6}$ & $(7.36\pm 0.15)\times 10^{-5}$ \\
\hline
$\phi$ & $1019.461\pm 0.016 \,$ & $4.249\pm 0.013$   & $(3.2 \pm 1.5) \times 10^{-8}$   & $(2.973\pm 0.034)\times 10^{-4}$ \\
\hline
$J/\psi$ &  $3096.900\pm 0.006$ & $(92.9\pm 2.8) \times 10^{-3}$ &  $ (3.87\pm 0.11) \times 10^{-5}$  & $(5.971\pm 0.032)\times 10^{-2}$ \\
\hline
$\psi(2S)$ &  $3686.10\pm 0.06$ & $(294\pm 8) \times 10^{-3}$ &  $(2.44\pm 0.30)\times 10^{-5}$ & $(7.93\pm 0.17)\times 10^{-3}$ \\
\hline
$\psi(3770)$ &  $3773.7\pm 0.3$ & $(27.2\pm 1) $ & -& $(9.6\pm 0.7)\times 10^{-6}$ \\
\hline
$\psi(4040)$ &  $4039\pm 1$ & $(80\pm 10)$ & -& $(1.07\pm 0.16)\times 10^{-5}$ \\
\hline
$\psi(4160)$ &  $4191\pm 5$ & $(70\pm 10) $ & -& $(6.9\pm 3.3)\times 10^{-6}$ \\
\hline
$\psi(4415)$ &  $4421\pm 4$ & $(62\pm 20)$ & -& $(9.4\pm 3.2)\times 10^{-6}$ \\
\hline
\end{tabular}
\caption{Experimental inputs for the determination of resonance contributions in $B\to \pi \ell^+\ell^-$, from Ref.~\cite{Zyla:2020zbs}.}
\label{tab:inputs}
\end{center}
\end{table*}

An estimate of $\eta_j$ can be obtained from the decay $\bar B_q\to P V_j\to P\ell^+\ell^-$. 
Focussing on the $P=\pi$ case, we can write
\begin{widetext}
\begin{align}
\begin{aligned}
 \mathcal{B}(B^{+} \to \pi^{+} V_{j}) \times \mathcal{B}(V_{j} \to \ell^+\ell^-) & \approx    \tau_{B^{+}} \frac{G_{F}^{2} \alpha^{2} \abs{ V_{tb}V^{\ast}_{t d}}^{2}}{1024 \pi^{5} m_{B}^{3}}  \int^{(m_{B}-m_{\pi})^{2}}_{4 m_{\ell}^{2}} d q^{2} \lambda^{3/2}(m_{B}^{2}, m_{\pi}^{2},q^{2}) 
\beta(q^{2})\left[ 1 - \frac{{\beta^2(q^{2})}}{3} \right] \times \\
& \times  (16 \pi^{2})^{2} \abs{A^{\mathrm{res}}_{j}(q^{2})}^{2}   \eta_{j}^{2} 
\times \left\{ 
\begin{array}{ll} 
\frac{q^{2}}{m_{j}^{2} } \left|   \frac{\labdc}{\labdt}  \right|^{2} & [ V_j = J/\psi, \psi(2S) ]~,\\
1  &   [ V_j = \rho, \omega, \phi ]~,
 \end{array}
 \right.
\end{aligned}
\label{eq:BtoVP}
\end{align}
\end{widetext}
where we have explicitly separated charmonia and light resonances.
In the narrow-width approximation (NWA) we have
\begin{align}
\abs{A_{j}^{\mathrm{res}}(q^{2})}^{2}  = \frac{m_{j}^{2} \Gamma_{j}^{2}}{(q^{2}-m_{j}^{2})^{2} +m_{j}^{2} \Gamma_{j}^{2}}\to m_{j} \Gamma_{j} \, \pi \, \delta(q^{2} - m_{j}^{2 })\,,
\end{align}
which, setting also $m_{\ell}=0$, allows us to further simplify (\ref{eq:BtoVP}) to 
\begin{align}
\begin{aligned}
&\mathcal{B}(B^{+} \to \pi^{+} V_{j}) \times \mathcal{B}(V_{j} \to \ell^+ \ell^-) \approx  \frac{    \ \tau_{B^{+}} G_{F}^{2} \alpha^{2}  }{6 m_{B}^{3} } \times \\
&\times \abs{ V_{tb}V^{\ast}_{t d}}^{2}  \lambda^{3/2}(m_{B}^{2}, m_{\pi}^{2},m_{j}^{2})\, \eta_{j}^{2}\, m_{j} \Gamma_{j}   \,.
\end{aligned}
\end{align}
Using this expression and  the inputs in~\Table{tab:inputs} we find the $\eta_{j}$ values reported 
in the second column of \Table{tab:etas}. We checked explicitly that relaxing the NWA by considering a variable width for broader resonances, like the $\rho$, does not affect our results significantly.
\subsubsection{Constraints on the charmonium states from \texorpdfstring{$B\to K \ell^+ \ell^-$}{B->Kll}}

By definition, the $Y$ functions in (\ref{eq:Ydef}) are process-dependent.  However, the 
$Y_{c\bar c}(q^2)$ function for $B\to \pi$  
is expected to be very close to the one for $B\to K$ decays, analyzed recently in Ref.~\cite{Cornella:2020aoq}. Due to the different CKM structure, in the latter case $Y_{c\bar c}(q^2)$ enters the decay rate via the combination
\begin{align}
\left. \mathcal{C}_9^{\ell,\text{eff}}(q^2) \right|_{b \to s} =  \left. \mathcal{C}_9^\ell \right|_{b \to s} +Y_{c\bar c}(q^2)~,
\end{align}
where we have used $\labsc \approx -  \labst$. 

The $Y_{c\bar c}(q^2)$ functions for $B\to \pi \ell^+\ell^-$ and $B\to K\ell^+\ell^-$ decays are expected to coincide in the limit where we neglect $SU(3)$-breaking effects in the subleading spectator-quark contributions. This expectation is well supported by the comparison $\mathcal{B}(B^{+} \to \pi^{+} V_{j})$ vs.~$\mathcal{B}(B^{+} \to K^{+} V_{j})$, which exhibits a universal scaling,
\begin{align}
 \frac{ \mathcal{B}(B^{+} \to \pi^{+} V_{j})  }{ \mathcal{B}(B^{+} \to K^{+} V_{j}) } \approx  \left|  \frac{ \labdc}{ \labsc} \right|^{2}
\approx 0.04\,,
\end{align} 
in the well-measured cases of the first two charmonium states. This allows us, within our $B\to \pi \ell^+ \ell^- $ analysis, to use the magnitudes $\eta_j$ and phases $\delta_j$ extracted from $B\to K\ell^+ \ell^- $ in  Ref.~\cite{Aaij:2016cbx} for all the charmonia, reducing the uncertainty significantly. Note that a similar argument cannot be made for the light resonances due to the different relative weight of the (different) light-quark operators in $b\to s$ and $b\to d$ transitions.

The applicability of $B\to K$ data from Ref.~\cite{Aaij:2016cbx} is even crucial for the charmonia above the $\psi (2S)$, where no $B\to \pi V_j$ data is available. The $B\to K V_j$ branching ratios reported in~\cite{Aaij:2016cbx} have a slight dependence on the sign of the $J/\psi$ and $\psi(2S)$ phases. However, this is  within uncertainties. For this reason we use a weighted average as reference value. Performing the CKM-rescaling illustrated above, we estimate
\begin{align}
\begin{aligned}
& 	\cB(B\to \pi \psi_j) \times \cB(\psi_j \to e^+e^-) =  \\
&	\qquad = \left\{ \begin{array}{ll}  
(8.16\pm1.54)\times 10^{-11}  	&  	\psi_j = \psi(3770)\,, 	\\
(2.26\pm0.75)\times 10^{-11} 	&  	\psi_j =\psi(4040)\,, 	\\
(1.43\pm 0.22)\times 10^{-10}	& 	\psi_j =\psi(4160)\,,	\\
(3.52\pm1.15)\times 10^{-11} 	&  	\psi_j =\psi(4415)\,,
\end{array} \right.
\end{aligned}
\end{align}
which yield the  $\eta_V$ reported in the last column of~\Table{tab:etas}.

\begin{table}
\begin{center}
\renewcommand{\arraystretch}{1.1} 
\begin{tabular}{|c || c | c|}
\hline
 $V$ & $\eta_V$~from $B\to \pi$  &   $\eta_V$~from $B\to K$  \\  \hline\hline
$\rho $  & $(1.02\pm 0.07)\times 10^{-2}$ & -\\   \hline
$\omega$ &  $(4.8\pm 0.2)\times 10^{-2}$ &-  \\   \hline
$\phi$   & $(8\pm2)\times 10^{-3}$ & - \\   \hline
$\psi$  & $(26.0\pm0.4)$ &  - \\  \hline
$\psi$(2S)   & $(5.6\pm0.4)$ &  -\\   \hline
$\psi(3770)$   & -  & $(1.3\pm0.1)\times10^{-2}$\\   \hline
$\psi(4040)$   & - & $(4.8\pm0.8)\times 10^{-3}$\\   \hline
$\psi(4160)$   & - &  $(1.5\pm0.1)\times10^{-2}$\\  \hline
$\psi(4415)$   &- & $(1.1\pm0.2)\times 10^{-2}$\\ \hline
\end{tabular}
\end{center}
\caption{Numerical results for the moduli of the $\eta_V$  parameters controlling long-distance effects in $B^+\to\pi^+\ell^+\ell^-$. The uncertainties include only the experimental errors  on the $\mathcal{B}(B^+\to\pi^+ V)$ and $\mathcal{B}(V\to e^+e^-)$ values used to determine a given $\eta_V$.}
\label{tab:etas}
\end{table}

\section{Numerical analysis}

\subsection{Dilepton spectrum}
\label{sect:NumA}

\begin{figure*}
\includegraphics[scale=0.45]{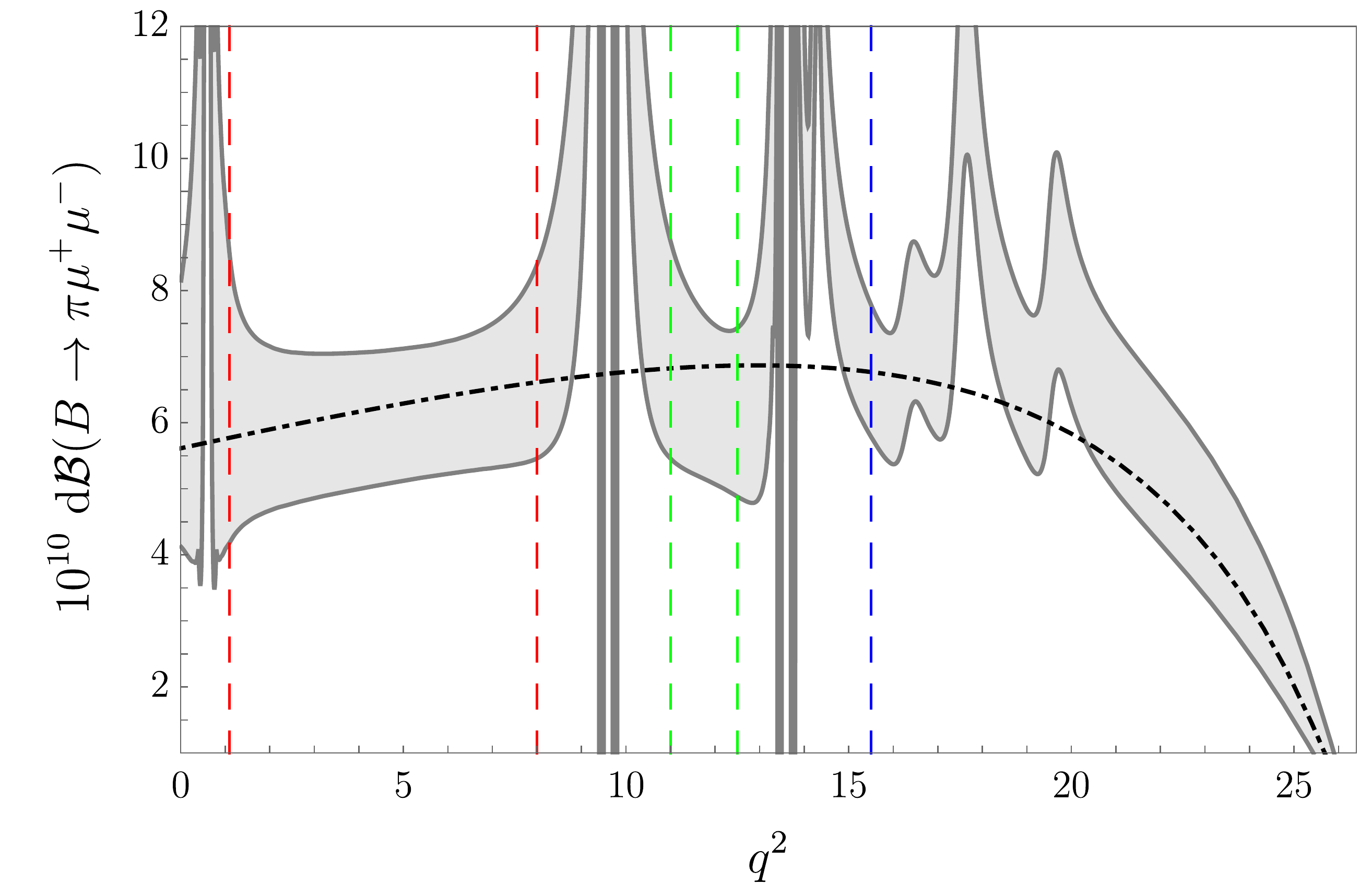}
\caption{Dilepton spectrum in $B^+\to \pi^+ \mu^+\mu^-$ within the SM. The dotted-dashed line indicate the perturbative contribution (ignoring its parametric uncertainty). The gray error band represents the 68\% interval after sampling over magnitudes and phases of the vector resonance contributions. The vertical dashed lines denote the three short-distance dominated regions where we provide precise estimates of $\Rpi$ (see text for details).}
\label{fig:spectrum}
\end{figure*}

Having discussed the general decomposition of the $B^+\to \pi^+ \ell^+\ell^-$ decay amplitude, we are ready to present numerical predictions within the SM and beyond. We begin with an analysis of the dilepton invariant mass spectrum in the SM to identify viable $q^2$-regions in which to perform precise LFU tests.

In \sec{sec:2} we discussed how to estimate the parameters  $\eta_V$, which control the magnitude of long-distance contributions, obtaining the results summarized in Table~\ref{tab:etas}. We also pointed out that, in the cases of the charmonia, we can use the results for the strong phases 
%(i.e.~the phases of the $\eta_V$) 
determined from $B^+\to K^+\mu^+\mu^-$ in Ref.~\cite{Aaij:2016cbx}. However, we have no constraints on the strong phases for the light-quark resonances, which we treat as  free parameters.

In \fig{fig:spectrum} we show the differential branching fraction for the $B^+\to\pi^+\mu^+\mu^-$ decay obtained adopting the following procedure:
\begin{itemize}
\item[\textit{i})] sampling the magnitudes $\eta_V$ using independent Gaussian distributions, with central values and standard deviations defined by the figures in Table~\ref{tab:etas};
\item[\textit{ii})] randomly sampling the light-quark phases in the interval $[0,2\pi)$; 
\item[\textit{iii})] randomly choosing one of the four possible sets of solutions in Ref.~\cite{Aaij:2016cbx} for $J/\psi$ and $\psi(2S)$ phases and, within that set, using a multi-dimensional Gaussian distribution according to the central values and errors. 
\end{itemize}
 The gray band represents the 68\% confidence interval resulting from the sampling. The dashed-dotted black line indicates the contribution obtained using only the perturbative value of  $\mathcal{C}_9$. We checked explicitly that the size of the uncertainties depends mainly on the lack of information on the strong phases and on the form factors uncertainties, while the errors 
 on the  $\eta_V$ have a subleading impact. We stress that the uncertainties related to the form factors cancel in the LFU ratio.
  
 From the plot in \fig{fig:spectrum} we identify three regions where the rate is dominated by perturbative contributions:  \textit{i})~the low-$q^2$ region, $q^2 \in [1.1,8]\,\text{GeV}^2$; \textit{ii})~the region between the two narrow charmonium states, $q^2 \in [11,12.5]\,\text{GeV}^2$,
 and \textit{iii})~the high-$q^2$ region,  $q^2 > 15.5\,\text{GeV}^2$. 

\subsection{The LFU ratio}
\label{sect:3B}

\begin{figure*}[t]
\begin{center}
\subfloat{\includegraphics[scale=0.58]{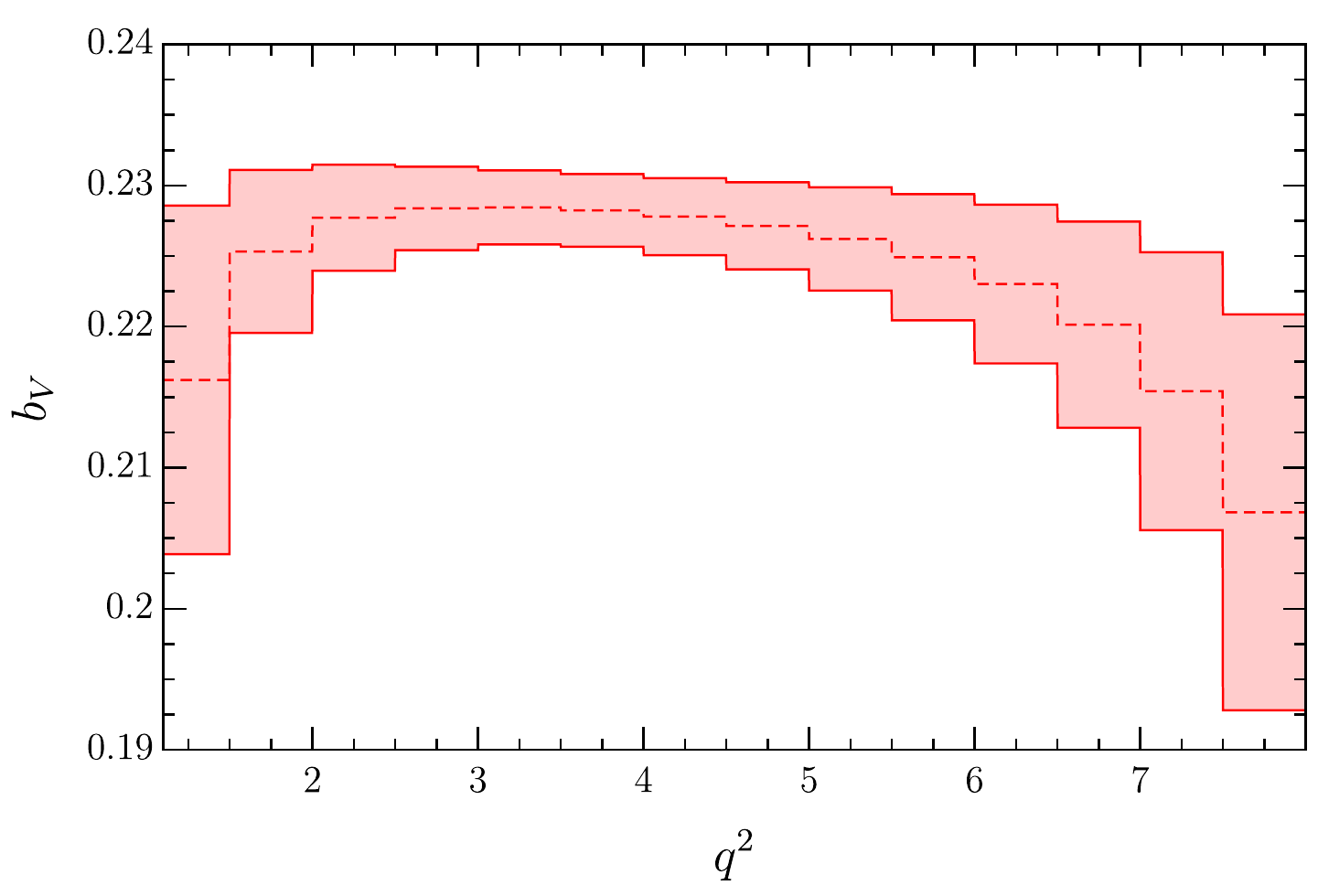}}
\hspace{4mm}
\subfloat{\includegraphics[scale=0.58]{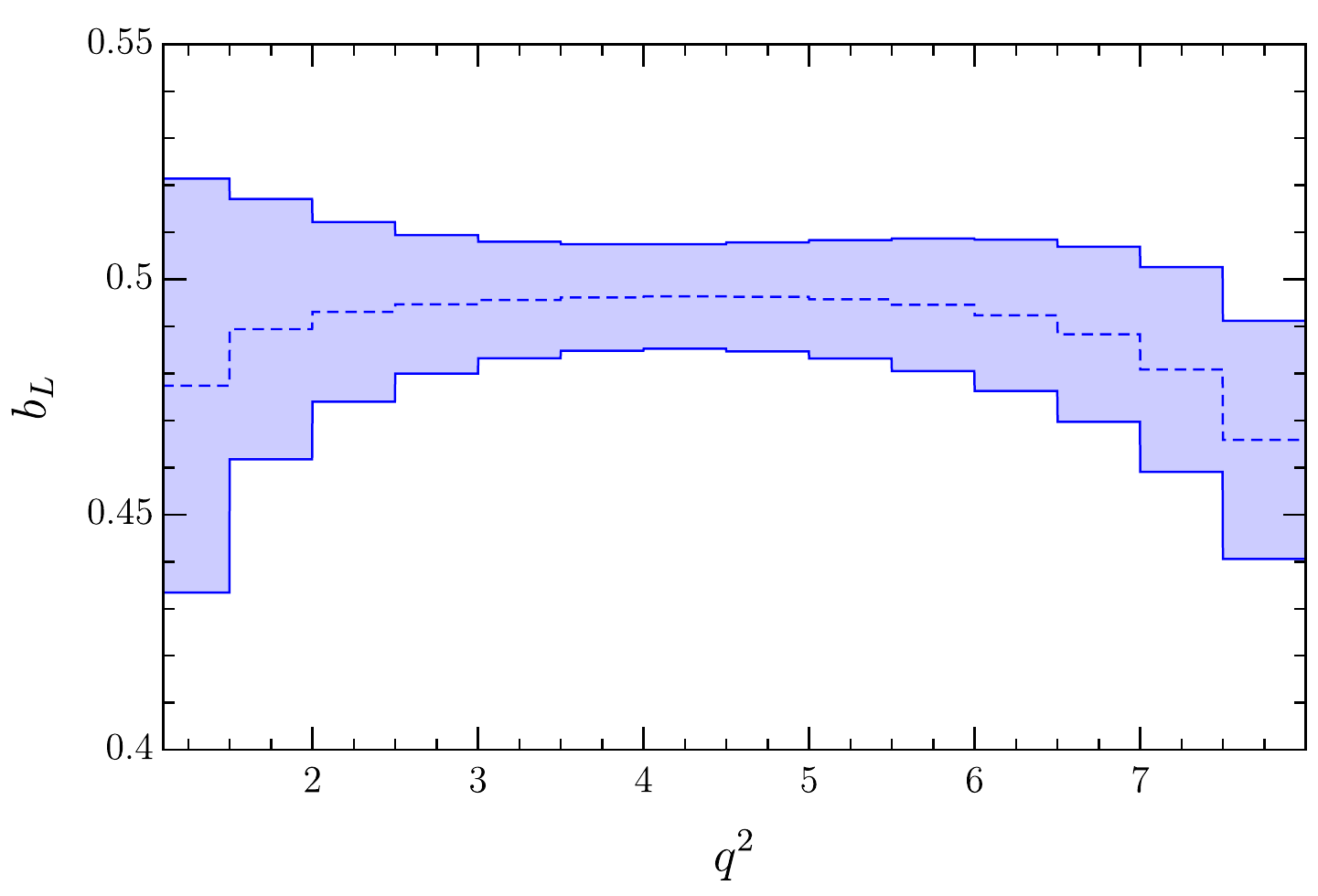}}
\end{center}
\caption{Predictions for the parameters $b_V$ (left) and $b_L$ (right) in small $q^2$ bins in the low $q^2$ region
(see main text). Numerical results for the different bins and their correlation matrix are given in \app{app:1}.}
\label{fig:binned_b}
\end{figure*}

We are now ready to analyze the LFU ratio $\Rpi$, defined in eq.~(\ref{eq:Rpidef}), in the three regions identified above. To estimate the sensitivity to the class of new physics (NP) models we are interested in, we modify the perturbative values of $\mathcal{C}^\ell_{9,10}$ as follows:
\begin{align}
\begin{aligned}
\mathcal{C}_{9,10}^e &= \left. \mathcal{C}_{9,10}^e \right|_{\rm SM}\,,  \\ 
\mathcal{C}_{9,10}^\mu &= \left. \mathcal{C}_{9,10}^\mu \right|_{\rm SM}  + \Delta \mathcal{C}_{9,10}\,.
\end{aligned}
\end{align}
This allows us to expand $\Rpi$ in the (small) NP contributions $\Delta \mathcal{C}_{9,10}$ as
\begin{align}
\begin{aligned}
\Rpi[q^2_\text{min}, q^2_\text{max}]\, =\,  &  \Rpi^{\rm SM} + b_V \Delta \mathcal{C}_9 +  (b_V -   b_L)  \Delta \mathcal{C}_{10} \\ 
& + \O\left( \Delta \mathcal{C}_9^2 ,\Delta \mathcal{C}_9  \Delta \mathcal{C}_{10} , \Delta \mathcal{C}_{10}^2 \right)\,,
\end{aligned}
\label{eq:RpiNP}
\end{align}
where $\Rpi^{\rm SM}$ and $b_{V,L}$ are adimensional numerical coefficients which, in general, 
depend on the $q^2$ region of interest. The definition of  $b_{V,L}$ is such that $b_{V}$ ($b_L$) 
controls the effect of a pure vectorial (left-handed) NP contribution in the lepton current.

Performing the same sampling discussed in Sect.~\ref{sect:NumA}, with non-vanishing $\Delta \mathcal{C}_{9,10}$, we estimated the numerical values of $\Rpi^{\rm SM}$ and $b_{V,L}$, as well as the size of the quadratic terms in eq.~\eqref{eq:RpiNP}. In all regions of interest we find 
\begin{align}
\left| \Rpi^{\rm SM} - 1 \right| < 0.01~.
\label{rpiSM}
\end{align}
This is not surprising since we have not included QED corrections and, in this limit, the only breaking of universality within the SM is due to tiny phase-space corrections. The result in eq.~\eqref{rpiSM} remains true also when QED effects are taken into account, provided $\Rpi^{\rm SM}$ is defined in a photon-inclusive way~\cite{Bordone:2016gaq,Isidori:2020acz}.\footnote{More precisely, lepton-dependent log-enhanced QED  corrections vanish only if we define $q^2$ from hadron momenta, i.e.  $q^2= (p_B - k)^2 \not= (p_{\ell^+} + p_{\ell^-})^2$~\cite{Isidori:2020acz}.}

As far as the NP coefficients are concerned, we first note that quadratic terms in $\Delta \mathcal{C}_{9,10}$ lead to corrections of at most $1\%$ in  $\Rpi$, i.e.~around or below the level of QED corrections, if $|\Delta \mathcal{C}_{9,10}| < 1.0$. This condition is what we expect in the most plausible NP scenarios. For instance, in models based on a minimally broken $U(2)^5$ symmetry, $b\to s\ell^+\ell^-$ data implies $\Delta \mathcal{C}_{9} = - \Delta \mathcal{C}_{10} = -0.43 \pm 0.11$~\cite{Fuentes-Martin:2019mun}.

Focusing on the linear terms only, and proceeding as in Sect.~\ref{sect:NumA}, we find the following results for the three regions of interest:
\begin{align}
\begin{aligned}
b_V=&\, 0.22\pm0.01 \,, \qquad&  q^2 \in &\, [1.1,8.0]\,\text{GeV}^2\,, \\
b_V=&\, 0.19\pm0.03  \,, \qquad&  q^2 \in &\, [11.0, 12.5]\,\text{GeV}^2\,, \\
b_V= &\,0.249\pm0.004\,, \qquad&  q^2 > &\, 15.5\,\text{GeV}^2\,, 
\end{aligned}
\label{eq:results_V}
\end{align}
and 
\begin{align}
\begin{aligned}
b_L=&\, 0.49\pm0.04 \,, \qquad&  q^2 \in &\, [1.1,8.0]\,\text{GeV}^2\,, \\ 
b_L=&\, 0.47\pm0.04 \,, \qquad&  q^2 \in &\, [11.0, 12.5]\,\text{GeV}^2\,, \\
b_L= &\,0.50\pm0.01 \,, \qquad&  q^2 > &\, 15.5\,\text{GeV}^2\,.
\end{aligned}
\label{eq:results_L}
\end{align}

In \fig{fig:binned_b} we show the results of a closer inspection of the low-$q^2$ region.
We divide the region into finer $q^2$ bins of width $0.5\, \text{GeV}^2$, except for the 
first bin, which is defined as $[1.1,1.5]~\text{GeV}^2$, and compute the $b_{V,L}$ for each bin. We find that the relative uncertainties span from $\sim 1\%$ to $\sim 10\%$, the largest value being for the bins closer to the resonances. The predictions are highly correlated, especially for neighboring bins, as is expected for a slowly-changing function. Central values, uncertainties and the full correlation matrices are reported in \app{app:1}.

The figures in eqs.~\eqref{eq:results_V} and~\eqref{eq:results_L} 
indicate that we can perform very precise tests of LFU in $B\to\pi \ell^+ \ell^-$ decays, irrespective of the sizable long-distance contributions affecting these modes. For instance, in the class of NP models analyzed in Ref.~\cite{Fuentes-Martin:2019mun}, we predict 
\begin{align}
\left. \Rpi \right|^{\rm NP}_{U(2)^5} = 0.79 \pm 0.05_{\rm NP} \pm
0.02_{\rm LD}\,,
\label{eq:RpiNPres}
\end{align}
in the low-$q^2$ region.
Here the first error is due to the NP model (or better the NP Wilson coefficients extracted from $b\to s$ data),  while the second (subleading) error is due to long-distance effects.

Experimental prospects for the measurement of $R_\pi$ in the various phases of the LHCb Upgrade II are listed in Ref.~\cite{Bediaga:2018lhg}. Considering only the $[1.1,6.0]~\text{GeV}^2$ bin, and using the prediction in eq.~\eqref{eq:RpiNPres} as a reference value, we should expect to observe a deviation from the SM (i.e.~$R_\pi \not=1$) exceeding the $\sim 3\sigma$ level only with $300\, \text{fb}^{-1}$. However, as we have shown, the low-$q^2$ bin can  be extended up to $8.0~\text{GeV}^2$ with a significant statistical gain.  Moreover, also the  high-$q^2$ region is theoretically clean. Extending the measurement of $R_\pi$ in both these directions, a similar level of significance could be reached already  with $50\, \text{fb}^{-1}$.

\section{CONCLUSIONS}

 The evidence of LFU violation, accompanied by the other anomalies observed in semi-leptonic $B$ decays, may represent the first hint of physics beyond the SM. While it is premature to draw conclusions, the pattern of anomalies is tantalizingly coherent and, when combined, consistently points towards new dynamics of common short-distance origin.
 
 In order to understand the flavor structure of such new dynamics, it is very important 
to collect additional indications of LFU violation in other low-energy processes. As we have shown in this letter, one candidate is the $R_\pi$ ratio defined in eq.~(\ref{eq:Rpidef}). By means of a general analysis of long-distance contributions in $B\to \pi \ell^+\ell^-$ decays, we have shown that $R_\pi$ can be predicted with high accuracy, both within and beyond the SM, in large regions of the dilepton invariant mass spectrum. The numerical coefficients reported in
\sec{sect:3B} allow for the extraction of precise short-distance information from this observable. 
The combination of future data on $R_\pi$ and $R_{K^{(*)}}$ would be an extremely valuable tool to determine the orientation of the new dynamics in quark-flavor space, possibly confirming the link between
LFU anomalies and Yukawa hierarchies.

\section*{Acknowledgements}
This project has received funding from the European Research Council (ERC) under the European Union’s Horizon 2020 research and innovation programme under grant agreement 833280 (FLAY), and by the Swiss Na- tional Science Foundation (SNF) under contract 200021-175940.
The work of M.B.  is supported by Deutsche Forschungsgemeinschaft (DFG, German Research Foundation) under grant 396021762 - TRR 257 ``Particle Physics Phenomenology after the Higgs Discovery'' and by the Italian Ministry of Research (MIUR) under grant PRIN 20172LNEEZ.
M.K. gratefully acknowledges funding by the Deutsche Forschungsgemeinschaft (DFG, German Research Foundation) – Project-ID 196253076 – TRR 110.

\appendix 
\section{Correlations for the binned prediction}
\label{app:1}
In \Table{tab:binned_b} we report the central values and uncertainties on the quantities $b_V$ and $b_L$ in sixteen $q^2$ bins. The first fourteen bins range from $1.1\,\GeV^2$ to $8.0\,\GeV^2$, where the first bin is defined as  $q^2  \in[1.1,1.5]\,\GeV^2$, and the following thirteen bins have equal size of $0.5\, \GeV^2$ up to $q^2 = 8.0\, \GeV^2$. Additionally, we include the bins $q^2 \in [11,12.5]\,\GeV^2 $ and $q^2 \in [15.5,26.4]\,\GeV^2$. The correlations are given in Tables~\ref{tab:binned_bV_corr} and~\ref{tab:binned_bL_corr}.

\begin{table}[t]
\begin{center}
\renewcommand{\arraystretch}{1.4} 
\begin{tabular}{|c||c|c|}
\hline
 \parbox{0.11\textwidth}{bin [GeV$^2$]}  & \parbox{0.11\textwidth}{$b_V$} &   \parbox{0.11\textwidth}{$b_L$}  \\ \hline
$[1.1, 1.5] $  &  $0.2162(124)$    &  $0.4774(440)$ \\
$[1.5, 2.0]$ &  $0.2253(58)$    &   $0.4894(277)$ \\
$[2.0, 2.5]$   & $0.2277(38)$     &   $0.4931(191)$ \\
$[2.5, 3.0]$  & $0.2284(30)$     &    $0.4947(147)$\\
$[3.0, 3.5]$   & $0.2284(26)$     &   $0.4956(124)$ \\
$[3.5, 4.0]$   &  $0.2282(26)$     &    $0.4962(113)$ \\
$[4.0, 4.5]$   & $0.2278(27)$     &   $0.4964(111)$ \\
$[4.5, 5.0]$   &  $0.2271(31)$     &   $0.4963(116)$ \\
$[5.0, 5.5]$   & $0.2262(37)$     &   $0.4958(126)$ \\
$[5.5, 6.0]$   & $0.2249(45)$     &   $0.4946(141)$ \\
$[6.0, 6.5]$   &  $0.2230(56)$     &    $0.4924(161)$\\
$[6.5, 7.0]$   & $0.2201(73)$     &    $0.4883(186)$\\
$[7.0, 7.5]$   &  $0.2154(98)$     &    $0.4809(218)$\\
$[7.5, 8.0]$   & $0.2068(140)$     &    $0.4659(253)$\\ \hline
$[11.0, 12.5]$   &  $0.1924(345)$     &   $0.4748(386)$  \\
$[15.5, 26.4]$   & $0.2490(39)$     &    $0.5044(72)$\\ \hline
\end{tabular}
\caption{Central values and uncertainties for the  $b_V$ and $b_L$ parameters 
in all $q^2$ bins.}
\label{tab:binned_b}
\end{center}
\end{table}

\begin{table*}
\renewcommand{\arraystretch}{1.1}
 \resizebox{1\textwidth}{!}{  
\begin{tabular}{ c c c c c c c c c c c c c c c c c  }
\toprule
$1$ & $0.9745$ & $0.9291$ & $0.8677$ & $0.7831$ & $0.6761$ & $0.5592$ & $0.4483$ & $0.3534$ & $0.2770$ & $0.2175$ & $0.1717$ & $0.1367$ & $0.1099$ & $-0.0045$ & $-0.0304$ \\
$0.9745$ & $1$ & $0.9856$ & $0.9452$ & $0.8749$ & $0.7755$ & $0.6601$ & $0.5464$ & $0.4465$ & $0.3645$ & $0.2996$ & $0.2490$ & $0.2098$ & $0.1796$ & $-0.0375$ & $-0.0694$ \\
$0.9292$ & $0.9856$ & $1$ & $0.9858$ & $0.9391$ & $0.8591$ & $0.7573$ & $0.6515$ & $0.5552$ & $0.4742$ & $0.4087$ & $0.3569$ & $0.3163$ & $0.2848$ & $-0.0990$ & $-0.1281$ \\
$0.8677$ & $0.9452$ & $0.9858$ & $1$ & $0.9830$ & $0.9308$ & $0.8515$ & $0.7617$ & $0.6758$ & $0.6010$ & $0.5390$ & $0.4890$ & $0.4494$ & $0.4185$ & $-0.1861$ & $-0.1988$ \\
$0.7831$ & $0.8749$ & $0.9391$ & $0.9830$ & $1$ & $0.9818$ & $0.9323$ & $0.8655$ & $0.7959$ & $0.7321$ & $0.6775$ & $0.6323$ & $0.5958$ & $0.5672$ & $-0.2920$ & $-0.2718$ \\
$0.6761$ & $0.7755$ & $0.8591$ & $0.9308$ & $0.9818$ & $1$ & $0.9839$ & $0.9443$ & $0.8952$ & $0.8461$ & $0.8018$ & $0.7639$ & $0.7327$ & $0.7079$ & $-0.4018$ & $-0.3343$ \\
$0.5592$ & $0.6601$ & $0.7573$ & $0.8515$ & $0.9323$ & $0.9839$ & $1$ & $0.9879$ & $0.9602$ & $0.9272$ & $0.8946$ & $0.8653$ & $0.8404$ & $0.8202$ & $-0.4997$ & $-0.3767$ \\
$0.4483$ & $0.5464$ & $0.6515$ & $0.7617$ & $0.8655$ & $0.9443$ & $0.9880$ & $1$ & $0.9919$ & $0.9739$ & $0.9527$ & $0.9319$ & $0.9133$ & $0.8978$ & $-0.5774$ & $-0.3982$ \\
$0.3534$ & $0.4465$ & $0.5552$ & $0.6758$ & $0.7959$ & $0.8952$ & $0.9602$ & $0.9919$ & $1$ & $0.9948$ & $0.9835$ & $0.9702$ & $0.9573$ & $0.9459$ & $-0.6354$ & $-0.4034$ \\
$0.2770$ & $0.3645$ & $0.4742$ & $0.6010$ & $0.7321$ & $0.8461$ & $0.9272$ & $0.9739$ & $0.9948$ & $1$ & $0.9968$ & $0.9898$ & $0.9816$ & $0.9735$ & $-0.6777$ & $-0.3988$ \\
$0.2175$ & $0.2996$ & $0.4087$ & $0.5390$ & $0.6775$ & $0.8018$ & $0.8946$ & $0.9527$ & $0.9835$ & $0.9968$ & $1$ & $0.9980$ & $0.9937$ & $0.9884$ & $-0.7086$ & $-0.3895$ \\
$0.1717$ & $0.2490$ & $0.3569$ & $0.4890$ & $0.6323$ & $0.7639$ & $0.8653$ & $0.9319$ & $0.9703$ & $0.9898$ & $0.9980$ & $1$ & $0.9988$ & $0.9958$ & $-0.7308$ & $-0.3798$ \\
$0.1367$ & $0.2098$ & $0.3163$ & $0.4494$ & $0.5958$ & $0.7327$ & $0.8404$ & $0.9133$ & $0.9573$ & $0.9816$ & $0.9937$ & $0.9988$ & $1$ & $0.9990$ & $-0.7453$ & $-0.3738$ \\
$0.1099$ & $0.17958$ & $0.2848$ & $0.4185$ & $0.5672$ & $0.7079$ & $0.8202$ & $0.8978$ & $0.9459$ & $0.9735$ & $0.9884$ & $0.9958$ & $0.9990$ & $1$ & $-0.7508$ & $-0.3773$ \\
$-0.0045$ & $-0.0375$ & $-0.0990$ & $-0.1861$ & $-0.2920$ & $-0.4018$ & $-0.4997$ & $-0.5774$ & $-0.6354$ & $-0.6777$ & $-0.7086$ & $-0.7308$ & $-0.7453$ & $-0.7508$ & $1$ & $-0.2306$ \\
$-0.0304$ & $-0.0694$ & $-0.1281$ & $-0.1988$ & $-0.2718$ & $-0.3343$ & $-0.3767$ & $-0.3982$ & $-0.4034$ & $-0.3988$ & $-0.3895$ & $-0.3798$ & $-0.3738$ & $-0.3773$ & $-0.2306$ & $1$ \\
  \toprule
\end{tabular}
}
\caption{Correlation matrix for the binned predictions of the parameter $b_V$.}
\label{tab:binned_bV_corr}
\end{table*}

\begin{table*}
\renewcommand{\arraystretch}{1.1}
 \resizebox{1\textwidth}{!}{  
\begin{tabular}{ c c c c c c c c c c c c c c c c c  }
\toprule
 $1$ & $0.9890$ & $0.9694$ & $0.9337$ & $0.8705$ & $0.7765$ & $0.6610$ & $0.5413$ & $0.4317$ & $0.3388$ & $0.2632$ & $0.2024$ & $0.1535$ & $0.1132$ & $0.0018$ & $0.0360$\\
 $0.9890$ & $1$ & $0.9934$ & $0.9687$ & $0.9158$ & $0.8305$ & $0.7212$ & $0.6051$ & $0.4970$ & $0.4042$ & $0.3278$ & $0.2658$ & $0.2152$ & $0.1724$ & $-0.0181$ & $0.0366$ \\
 $0.9694$ & $0.9934$ & $1$ & $0.9904$ & $0.9543$ & $0.8854$ & $0.7903$ & $0.6847$ & $0.5836$ & $0.4950$ & $0.4209$ & $0.3597$ & $0.3086$ & $0.2636$ & $-0.0535$ & $0.0356$ \\
 $0.9337$ & $0.9687$ & $0.9904$ & $1$ & $0.9863$ & $0.9407$ & $0.8664$ & $0.7772$ & $0.6878$ & $0.6070$ & $0.5377$ & $0.4791$ & $0.4287$ & $0.3820$ & $-0.1040$ & $0.0337$ \\
 $0.8705$ & $0.9158$ & $0.9543$ & $0.9863$ & $1$ & $0.9837$ & $0.9367$ & $0.8698$ & $0.7971$ & $0.7280$ & $0.6666$ & $0.6131$ & $0.5652$ & $0.5180$ & $-0.1677$ & $0.0314$  \\
 $0.7765$ & $0.8305$ & $0.8854$ & $0.9407$ & $0.9837$ & $1$ & $0.9843$ & $0.9441$ & $0.8922$ & $0.8384$ &  $0.7879$ & $0.7418$ & $0.6985$ & $0.6529$ & $-0.2381$ & $0.0289$ \\
 $0.6610$ & $0.7212$ & $0.7903$ & $0.8664$ & $0.9367$ & $0.9843$ & $1$ & $0.9874$ & $0.9577$ & $0.9209$ &  $0.8830$ & $0.8460$ & $0.8091$ & $0.7672$ & $-0.3066$ &  $0.0268$  \\
 $0.5413$ & $0.6051$ & $0.6847$ & $0.7772$ & $ 0.8698$ & $0.9441$ & $0.9874$ & $1$ & $0.9911$ & $0.9707$ & $0.9454$ & $0.9181$ & $0.8884$ & $0.8516$ & $-0.3675$ & $0.0257$ \\
 $0.4317$ & $0.4970$ & $0.5836$ & $0.6878$ & $0.7971$ & $0.8922$ & $0.9577$ & $0.9911$ & $1$ & $0.9940$ & $0.9801$ & $0.9618$ & $0.9394$ & $0.9085$ & $-0.4200$ & $0.0260$ \\
 $0.3388$ & $0.4042$ & $0.4950$ & $0.6070$ & $0.7280$ & $0.8384$ & $0.9209$ & $0.9707$ & $0.9940$ & $1$ & $0.9959$ & $0.9856$ & $0.9700$ & $0.9452$ & $-0.4663$ & $0.0277$\\
 $0.2632$ & $0.3278$ & $0.4209$ & $0.5377$ & $0.6666$ & $0.7879$ & $0.8830$ & $0.9454$ & $0.9801$  & $0.9959$ & $1$ & $0.9968$ & $0.9876$ & $0.9690$ & $-0.5101$ & $0.0312$ \\
 $0.2024$ & $0.2658$ & $0.3597$ & $0.4791$ & $0.6131$ & $0.7418$ & $0.8460$ & $0.9181$ & $0.9618$ & $0.9856$ & $0.9968$ & $1$ & $0.9969$ & $0.9848$ & $-0.5551$ & $0.0369$\\
 $0.1535$ & $0.2152$ & $0.3086$ & $0.4287$ & $0.5652$ & $0.6985$ & $0.8091$ & $0.8884$ & $0.9394$ & $0.9700$ & $0.9876$ & $0.9969$ & $1$ & $0.9953$  & $-0.6059$ & $0.0457$ \\
 $0.1132$ & $0.1724$ & $0.2636$ & $0.3820$ & $0.5180$ & $0.6529$ & $0.7672$ & $0.8516$ & $0.9085$ & $0.9452$ & $0.9690$ & $0.9848$ & $0.9953$ & $1$ & $-0.6711$ & $0.0607$  \\
 $0.0018$ & $-0.0181$ & $-0.0535$ & $-0.1040$ & $-0.1677$ & $-0.2381$ & $-0.3066$ & $-0.3675$ & $-0.4200$ & $-0.4663$ & $-0.5101$ & $-0.5551$ & $-0.6059$ & $-0.6711$ & $1$ & $-0.1629$ \\
 $0.0360$ & $0.0366$ & $0.0356$ & $0.0337$ & $0.0314$ & $0.0289$ & $0.0268$ & $0.0257$ & $0.0260$ & $0.0277$ & $0.0312$ & $0.0369$ & $0.0457$ & $0.0607$ & $-0.1629$ & $1$\\
  \toprule 
\end{tabular}
}
\caption{Correlation matrix for the binned predictions of the parameter $b_L$.}
\label{tab:binned_bL_corr}
\end{table*}

\newpage

\bibliographystyle{my.bst}
\bibliography{paper}

\providecommand{\href}[2]{#2}\begingroup\raggedright\begin{thebibliography}{10}

\bibitem{Aaij:2014ora}
{\bf LHCb}, R.~Aaij {\em et al.,}
  \href{http://dx.doi.org/10.1103/PhysRevLett.113.151601}{{\em Phys. Rev.
  Lett.} {\bf 113} (2014)  151601}, \href{http://arxiv.org/abs/1406.6482}{{\tt
  arXiv:1406.6482 [hep-ex]}}.

\bibitem{Aaij:2017vbb}
{\bf LHCb}, R.~Aaij {\em et al.,}
  \href{http://dx.doi.org/10.1007/JHEP08(2017)055}{{\em JHEP} {\bf 08} (2017)
  055}, \href{http://arxiv.org/abs/1705.05802}{{\tt arXiv:1705.05802
  [hep-ex]}}.

\bibitem{Aaij:2019wad}
{\bf LHCb}, R.~Aaij {\em et al.,}
  \href{http://dx.doi.org/10.1103/PhysRevLett.122.191801}{{\em Phys. Rev.
  Lett.} {\bf 122} (2019) no.~19, 191801},
  \href{http://arxiv.org/abs/1903.09252}{{\tt arXiv:1903.09252 [hep-ex]}}.

\bibitem{Abdesselam:2019wac}
{\bf Belle}, A.~Abdesselam {\em et al.,}
  \href{http://arxiv.org/abs/1904.02440}{{\tt arXiv:1904.02440 [hep-ex]}}.

\bibitem{Hiller:2003js}
G.~Hiller and F.~Kruger,
  \href{http://dx.doi.org/10.1103/PhysRevD.69.074020}{{\em Phys. Rev. D} {\bf
  69} (2004)  074020}, \href{http://arxiv.org/abs/hep-ph/0310219}{{\tt
  arXiv:hep-ph/0310219}}.

\bibitem{Bordone:2016gaq}
M.~Bordone, G.~Isidori, and A.~Pattori,
  \href{http://dx.doi.org/10.1140/epjc/s10052-016-4274-7}{{\em Eur. Phys. J. C}
  {\bf 76} (2016) no.~8, 440}, \href{http://arxiv.org/abs/1605.07633}{{\tt
  arXiv:1605.07633 [hep-ph]}}.

\bibitem{Hiller:2014yaa}
G.~Hiller and M.~Schmaltz,
  \href{http://dx.doi.org/10.1103/PhysRevD.90.054014}{{\em Phys. Rev. D} {\bf
  90} (2014)  054014}, \href{http://arxiv.org/abs/1408.1627}{{\tt
  arXiv:1408.1627 [hep-ph]}}.

\bibitem{Aaij:2020nrf}
{\bf LHCb}, R.~Aaij {\em et al.,}
  \href{http://dx.doi.org/10.1103/PhysRevLett.125.011802}{{\em Phys. Rev.
  Lett.} {\bf 125} (2020) no.~1, 011802},
  \href{http://arxiv.org/abs/2003.04831}{{\tt arXiv:2003.04831 [hep-ex]}}.

\bibitem{Aaij:2015oid}
{\bf LHCb}, R.~Aaij {\em et al.,}
  \href{http://dx.doi.org/10.1007/JHEP02(2016)104}{{\em JHEP} {\bf 02} (2016)
  104}, \href{http://arxiv.org/abs/1512.04442}{{\tt arXiv:1512.04442
  [hep-ex]}}.

\bibitem{LHCb:2020zud}
{\bf ATLAS, CMS, LHCb} \href{http://arxiv.org/abs/LHCb-CONF-2020-002,
  CERN-LHCb-CONF-2020-002}{{\tt LHCb-CONF-2020-002, CERN-LHCb-CONF-2020-002}}.

\bibitem{Beneke:2017vpq}
M.~Beneke, C.~Bobeth, and R.~Szafron,
  \href{http://dx.doi.org/10.1103/PhysRevLett.120.011801}{{\em Phys. Rev.
  Lett.} {\bf 120} (2018) no.~1, 011801},
  \href{http://arxiv.org/abs/1708.09152}{{\tt arXiv:1708.09152 [hep-ph]}}.

\bibitem{Beneke:2019slt}
M.~Beneke, C.~Bobeth, and R.~Szafron,
  \href{http://dx.doi.org/10.1007/JHEP10(2019)232}{{\em JHEP} {\bf 10} (2019)
  232}, \href{http://arxiv.org/abs/1908.07011}{{\tt arXiv:1908.07011
  [hep-ph]}}.

\bibitem{Alguero:2019ptt}
M.~Alguer\'o, B.~Capdevila, A.~Crivellin, S.~Descotes-Genon, P.~Masjuan,
  J.~Matias, M.~Novoa~Brunet, and J.~Virto,
  \href{http://dx.doi.org/10.1140/epjc/s10052-019-7216-3}{{\em Eur. Phys. J. C}
  {\bf 79} (2019) no.~8, 714}, \href{http://arxiv.org/abs/1903.09578}{{\tt
  arXiv:1903.09578 [hep-ph]}}. [Addendum: Eur.Phys.J.C 80, 511 (2020)].

\bibitem{Aebischer:2019mlg}
J.~Aebischer, W.~Altmannshofer, D.~Guadagnoli, M.~Reboud, P.~Stangl, and D.~M.
  Straub, \href{http://dx.doi.org/10.1140/epjc/s10052-020-7817-x}{{\em Eur.
  Phys. J. C} {\bf 80} (2020) no.~3, 252},
  \href{http://arxiv.org/abs/1903.10434}{{\tt arXiv:1903.10434 [hep-ph]}}.

\bibitem{Ciuchini:2020gvn}
M.~Ciuchini, M.~Fedele, E.~Franco, A.~Paul, L.~Silvestrini, and M.~Valli,
  \href{http://arxiv.org/abs/2011.01212}{{\tt arXiv:2011.01212 [hep-ph]}}.

\bibitem{Aaij:2015yra}
{\bf LHCb}, R.~Aaij {\em et al.,}
  \href{http://dx.doi.org/10.1103/PhysRevLett.115.111803}{{\em Phys. Rev.
  Lett.} {\bf 115} (2015) no.~11, 111803},
  \href{http://arxiv.org/abs/1506.08614}{{\tt arXiv:1506.08614 [hep-ex]}}.
  [Erratum: Phys.Rev.Lett. 115, 159901 (2015)].

\bibitem{Lees:2013uzd}
{\bf BaBar}, J.~Lees {\em et al.,}
  \href{http://dx.doi.org/10.1103/PhysRevD.88.072012}{{\em Phys. Rev. D} {\bf
  88} (2013) no.~7, 072012}, \href{http://arxiv.org/abs/1303.0571}{{\tt
  arXiv:1303.0571 [hep-ex]}}.

\bibitem{Hirose:2016wfn}
{\bf Belle}, S.~Hirose {\em et al.,}
  \href{http://dx.doi.org/10.1103/PhysRevLett.118.211801}{{\em Phys. Rev.
  Lett.} {\bf 118} (2017) no.~21, 211801},
  \href{http://arxiv.org/abs/1612.00529}{{\tt arXiv:1612.00529 [hep-ex]}}.

\bibitem{Aaij:2017deq}
{\bf LHCb}, R.~Aaij {\em et al.,}
  \href{http://dx.doi.org/10.1103/PhysRevD.97.072013}{{\em Phys. Rev. D} {\bf
  97} (2018) no.~7, 072013}, \href{http://arxiv.org/abs/1711.02505}{{\tt
  arXiv:1711.02505 [hep-ex]}}.

\bibitem{Abdesselam:2019dgh}
{\bf Belle}, A.~Abdesselam {\em et al.,}
  \href{http://arxiv.org/abs/1904.08794}{{\tt arXiv:1904.08794 [hep-ex]}}.

\bibitem{Bhattacharya:2014wla}
B.~Bhattacharya, A.~Datta, D.~London, and S.~Shivashankara,
  \href{http://dx.doi.org/10.1016/j.physletb.2015.02.011}{{\em Phys. Lett. B}
  {\bf 742} (2015)  370--374}, \href{http://arxiv.org/abs/1412.7164}{{\tt
  arXiv:1412.7164 [hep-ph]}}.

\bibitem{Alonso:2015sja}
R.~Alonso, B.~Grinstein, and J.~Martin~Camalich,
  \href{http://dx.doi.org/10.1007/JHEP10(2015)184}{{\em JHEP} {\bf 10} (2015)
  184}, \href{http://arxiv.org/abs/1505.05164}{{\tt arXiv:1505.05164
  [hep-ph]}}.

\bibitem{Greljo:2015mma}
A.~Greljo, G.~Isidori, and D.~Marzocca,
  \href{http://dx.doi.org/10.1007/JHEP07(2015)142}{{\em JHEP} {\bf 07} (2015)
  142}, \href{http://arxiv.org/abs/1506.01705}{{\tt arXiv:1506.01705
  [hep-ph]}}.

\bibitem{Calibbi:2015kma}
L.~Calibbi, A.~Crivellin, and T.~Ota,
  \href{http://dx.doi.org/10.1103/PhysRevLett.115.181801}{{\em Phys. Rev.
  Lett.} {\bf 115} (2015)  181801}, \href{http://arxiv.org/abs/1506.02661}{{\tt
  arXiv:1506.02661 [hep-ph]}}.

\bibitem{Barbieri:2015yvd}
R.~Barbieri, G.~Isidori, A.~Pattori, and F.~Senia,
  \href{http://dx.doi.org/10.1140/epjc/s10052-016-3905-3}{{\em Eur. Phys. J. C}
  {\bf 76} (2016) no.~2, 67}, \href{http://arxiv.org/abs/1512.01560}{{\tt
  arXiv:1512.01560 [hep-ph]}}.

\bibitem{Buttazzo:2017ixm}
D.~Buttazzo, A.~Greljo, G.~Isidori, and D.~Marzocca,
  \href{http://dx.doi.org/10.1007/JHEP11(2017)044}{{\em JHEP} {\bf 11} (2017)
  044}, \href{http://arxiv.org/abs/1706.07808}{{\tt arXiv:1706.07808
  [hep-ph]}}.

\bibitem{Bordone:2017bld}
M.~Bordone, C.~Cornella, J.~Fuentes-Martin, and G.~Isidori,
  \href{http://dx.doi.org/10.1016/j.physletb.2018.02.011}{{\em Phys. Lett. B}
  {\bf 779} (2018)  317--323}, \href{http://arxiv.org/abs/1712.01368}{{\tt
  arXiv:1712.01368 [hep-ph]}}.

\bibitem{Fuentes-Martin:2020pww}
J.~Fuentes-Martin, G.~Isidori, J.~Pag\`es, and B.~A. Stefanek,
  \href{http://arxiv.org/abs/2012.10492}{{\tt arXiv:2012.10492 [hep-ph]}}.

\bibitem{Barbieri:2011ci}
R.~Barbieri, G.~Isidori, J.~Jones-Perez, P.~Lodone, and D.~M. Straub,
  \href{http://dx.doi.org/10.1140/epjc/s10052-011-1725-z}{{\em Eur. Phys. J. C}
  {\bf 71} (2011)  1725}, \href{http://arxiv.org/abs/1105.2296}{{\tt
  arXiv:1105.2296 [hep-ph]}}.

\bibitem{Blankenburg:2012nx}
G.~Blankenburg, G.~Isidori, and J.~Jones-Perez,
  \href{http://dx.doi.org/10.1140/epjc/s10052-012-2126-7}{{\em Eur. Phys. J. C}
  {\bf 72} (2012)  2126}, \href{http://arxiv.org/abs/1204.0688}{{\tt
  arXiv:1204.0688 [hep-ph]}}.

\bibitem{Barbieri:2012uh}
R.~Barbieri, D.~Buttazzo, F.~Sala, and D.~M. Straub,
  \href{http://dx.doi.org/10.1007/JHEP07(2012)181}{{\em JHEP} {\bf 07} (2012)
  181}, \href{http://arxiv.org/abs/1203.4218}{{\tt arXiv:1203.4218 [hep-ph]}}.

\bibitem{Greljo:2018tuh}
A.~Greljo and B.~A. Stefanek,
  \href{http://dx.doi.org/10.1016/j.physletb.2018.05.033}{{\em Phys. Lett. B}
  {\bf 782} (2018)  131--138}, \href{http://arxiv.org/abs/1802.04274}{{\tt
  arXiv:1802.04274 [hep-ph]}}.

\bibitem{DiLuzio:2018zxy}
L.~Di~Luzio, J.~Fuentes-Martin, A.~Greljo, M.~Nardecchia, and S.~Renner,
  \href{http://dx.doi.org/10.1007/JHEP11(2018)081}{{\em JHEP} {\bf 11} (2018)
  081}, \href{http://arxiv.org/abs/1808.00942}{{\tt arXiv:1808.00942
  [hep-ph]}}.

\bibitem{Cornella:2019hct}
C.~Cornella, J.~Fuentes-Martin, and G.~Isidori,
  \href{http://dx.doi.org/10.1007/JHEP07(2019)168}{{\em JHEP} {\bf 07} (2019)
  168}, \href{http://arxiv.org/abs/1903.11517}{{\tt arXiv:1903.11517
  [hep-ph]}}.

\bibitem{Fuentes-Martin:2019ign}
J.~Fuentes-Mart\'\i{}n, G.~Isidori, M.~K\"onig, and N.~Selimovi\'c,
  \href{http://dx.doi.org/10.1103/PhysRevD.101.035024}{{\em Phys. Rev. D} {\bf
  101} (2020) no.~3, 035024}, \href{http://arxiv.org/abs/1910.13474}{{\tt
  arXiv:1910.13474 [hep-ph]}}.

\bibitem{Fuentes-Martin:2019mun}
J.~Fuentes-Mart\'\i{}n, G.~Isidori, J.~Pag\`es, and K.~Yamamoto,
  \href{http://dx.doi.org/10.1016/j.physletb.2019.135080}{{\em Phys. Lett. B}
  {\bf 800} (2020)  135080}, \href{http://arxiv.org/abs/1909.02519}{{\tt
  arXiv:1909.02519 [hep-ph]}}.

\bibitem{Aaij:2015nea}
{\bf LHCb}, R.~Aaij {\em et al.,}
  \href{http://dx.doi.org/10.1007/JHEP10(2015)034}{{\em JHEP} {\bf 10} (2015)
  034}, \href{http://arxiv.org/abs/1509.00414}{{\tt arXiv:1509.00414
  [hep-ex]}}.

\bibitem{Ali:2013zfa}
A.~Ali, A.~Y. Parkhomenko, and A.~V. Rusov,
  \href{http://dx.doi.org/10.1103/PhysRevD.89.094021}{{\em Phys. Rev. D} {\bf
  89} (2014) no.~9, 094021}, \href{http://arxiv.org/abs/1312.2523}{{\tt
  arXiv:1312.2523 [hep-ph]}}.

\bibitem{Hou:2014dza}
W.-S. Hou, M.~Kohda, and F.~Xu,
  \href{http://dx.doi.org/10.1103/PhysRevD.90.013002}{{\em Phys. Rev. D} {\bf
  90} (2014) no.~1, 013002}, \href{http://arxiv.org/abs/1403.7410}{{\tt
  arXiv:1403.7410 [hep-ph]}}.

\bibitem{Hambrock:2015wka}
C.~Hambrock, A.~Khodjamirian, and A.~Rusov,
  \href{http://dx.doi.org/10.1103/PhysRevD.92.074020}{{\em Phys. Rev. D} {\bf
  92} (2015) no.~7, 074020}, \href{http://arxiv.org/abs/1506.07760}{{\tt
  arXiv:1506.07760 [hep-ph]}}.

\bibitem{Khodjamirian:2017fxg}
A.~Khodjamirian and A.~V. Rusov,
  \href{http://dx.doi.org/10.1007/JHEP08(2017)112}{{\em JHEP} {\bf 08} (2017)
  112}, \href{http://arxiv.org/abs/1703.04765}{{\tt arXiv:1703.04765
  [hep-ph]}}.

\bibitem{Kruger:1996cv}
F.~Kruger and L.~Sehgal,
  \href{http://dx.doi.org/10.1016/0370-2693(96)00413-3}{{\em Phys. Lett. B}
  {\bf 380} (1996)  199--204}, \href{http://arxiv.org/abs/hep-ph/9603237}{{\tt
  arXiv:hep-ph/9603237}}.

\bibitem{Khodjamirian:2012rm}
A.~Khodjamirian, T.~Mannel, and Y.~Wang,
  \href{http://dx.doi.org/10.1007/JHEP02(2013)010}{{\em JHEP} {\bf 02} (2013)
  010}, \href{http://arxiv.org/abs/1211.0234}{{\tt arXiv:1211.0234 [hep-ph]}}.

\bibitem{Lyon:2014hpa}
J.~Lyon and R.~Zwicky, \href{http://arxiv.org/abs/1406.0566}{{\tt
  arXiv:1406.0566 [hep-ph]}}.

\bibitem{Blake:2017fyh}
T.~Blake, U.~Egede, P.~Owen, K.~A. Petridis, and G.~Pomery,
  \href{http://dx.doi.org/10.1140/epjc/s10052-018-5937-3}{{\em Eur. Phys. J. C}
  {\bf 78} (2018) no.~6, 453}, \href{http://arxiv.org/abs/1709.03921}{{\tt
  arXiv:1709.03921 [hep-ph]}}.

\bibitem{Bobeth:2017vxj}
C.~Bobeth, M.~Chrzaszcz, D.~van Dyk, and J.~Virto,
  \href{http://dx.doi.org/10.1140/epjc/s10052-018-5918-6}{{\em Eur. Phys. J. C}
  {\bf 78} (2018) no.~6, 451}, \href{http://arxiv.org/abs/1707.07305}{{\tt
  arXiv:1707.07305 [hep-ph]}}.

\bibitem{Cornella:2020aoq}
C.~Cornella, G.~Isidori, M.~K\"onig, S.~Liechti, P.~Owen, and N.~Serra,
  \href{http://dx.doi.org/10.1140/epjc/s10052-020-08674-5}{{\em Eur. Phys. J.
  C} {\bf 80} (2020) no.~12, 1095}, \href{http://arxiv.org/abs/2001.04470}{{\tt
  arXiv:2001.04470 [hep-ph]}}.

\bibitem{Gorbahn:2004my}
M.~Gorbahn and U.~Haisch,
  \href{http://dx.doi.org/10.1016/j.nuclphysb.2005.01.047}{{\em Nucl. Phys. B}
  {\bf 713} (2005)  291--332}, \href{http://arxiv.org/abs/hep-ph/0411071}{{\tt
  arXiv:hep-ph/0411071}}.

\bibitem{Gubernari:2018wyi}
N.~Gubernari, A.~Kokulu, and D.~van Dyk,
  \href{http://dx.doi.org/10.1007/JHEP01(2019)150}{{\em JHEP} {\bf 01} (2019)
  150}, \href{http://arxiv.org/abs/1811.00983}{{\tt arXiv:1811.00983
  [hep-ph]}}.

\bibitem{Lattice:2015tia}
{\bf Fermilab Lattice, MILC}, J.~A. Bailey {\em et al.,}
  \href{http://dx.doi.org/10.1103/PhysRevD.92.014024}{{\em Phys. Rev. D} {\bf
  92} (2015) no.~1, 014024}, \href{http://arxiv.org/abs/1503.07839}{{\tt
  arXiv:1503.07839 [hep-lat]}}.

\bibitem{Zyla:2020zbs}
{\bf Particle Data Group}, P.~Zyla {\em et al.,}
  \href{http://dx.doi.org/10.1093/ptep/ptaa104}{{\em PTEP} {\bf 2020} (2020)
  no.~8, 083C01}.

\bibitem{Aaij:2016cbx}
{\bf LHCb}, R.~Aaij {\em et al.,}
  \href{http://dx.doi.org/10.1140/epjc/s10052-017-4703-2}{{\em Eur. Phys. J. C}
  {\bf 77} (2017) no.~3, 161}, \href{http://arxiv.org/abs/1612.06764}{{\tt
  arXiv:1612.06764 [hep-ex]}}.

\bibitem{Isidori:2020acz}
G.~Isidori, S.~Nabeebaccus, and R.~Zwicky,
  \href{http://dx.doi.org/10.1007/JHEP12(2020)104}{{\em JHEP} {\bf 12} (2020)
  104}, \href{http://arxiv.org/abs/2009.00929}{{\tt arXiv:2009.00929
  [hep-ph]}}.

\bibitem{Bediaga:2018lhg}
{\bf LHCb}, R.~Aaij {\em et al.,} \href{http://arxiv.org/abs/1808.08865}{{\tt
  arXiv:1808.08865 [hep-ex]}}.

\end{thebibliography}\endgroup

\end{document}